\documentclass[aps,prb,showpacs,preprint,amssymb]{revtex4}
\usepackage{amsmath}
\usepackage{lmodern}
\usepackage[T1]{fontenc}
\usepackage[cp1250]{inputenc}
\usepackage[polish,english]{babel}
\usepackage{graphicx}
\usepackage[mathscr]{euscript}
\usepackage{bm}
\newcommand{\be}{\begin{equation}}
\newcommand{\ee}{\end{equation}}
\newcommand{\ba}{\begin{eqnarray}}
\newcommand{\ea}{\end{eqnarray}}
\newcommand{\kB}{k_{\sss B}}
\newcommand{\di}{{\rm d}}
\newcommand{\cF}{{\cal F}}
\newcommand{\sss}{\scriptscriptstyle\mrm}
\newcommand{\kBT}{{k_{\sss{B}}T}}
\newcommand{\mrm}{\mathrm}
\newcommand{\TD}{T_\mrm{D}}
\newcommand{\qD}{{q_{\sss{D}}}}
\newcommand{\Pph}{P^\mrm{ph}}
\newcommand{\Sph}{S^\mrm{ph}}
\newcommand{\Wph}{W^\mrm{ph}}
\newcommand{\Pimp}{P^\mrm{imp}}
\newcommand{\Cimp}{C^\mrm{imp}}
\newcommand{\cimp}{c^\mrm{imp}}
\newcommand{\Wimp}{W^\mrm{imp}}
\newcommand{\Win}{W^\mrm{in}}
\newcommand{\Wmag}{W^\mrm{mag}}
\newcommand{\Wel}{W^\mrm{el}}
\newcommand{\Cmag}{C^\mrm{mag}}
\newcommand{\cmag}{c^\mrm{mag}}
\newcommand{\gmag}{g^\mrm{mag}}
\newcommand{\gimp}{g^\mrm{imp}}
\newcommand{\Pmag}{P^\mrm{mag}}
\newcommand{\rhomag}{\rho^\mrm{mag}}
\newcommand{\rhores}{\rho^\mrm{res}}
\newcommand{\rhoimp}{\rho^\mrm{imp}}
\newcommand{\rhospd}{\rho^\mrm{spd}}
\newcommand{\rhoph}{\rho^\mrm{ph}}
\newcommand{\veps}{{\vep}_s}
\newcommand{\vepF}{\vep_{\sss{F}}}
\newcommand{\jex}{{j_\mrm{ex}}}
\newcommand{\vepk}{\vep_{\bk}}
\newcommand{\ovl}{\overline}
\newcommand{\pt}{\partial}
\newcommand{\mbf}{\mathbf}
\newcommand{\vep}{\varepsilon}
\newcommand{\kF}{\mbf{k}_{\sss{F}}}         
\newcommand{\cJ}{{\cal J}}
\newcommand{\bk}{{\bm{k}}}
\newcommand{\bv}{{\bm{v}}}
\newcommand{\bu}{{\bm{u}}}
\begin{document}
\title{Crystal field influence on the thermal conductivity of the rare-earth metallic paramagnets}

\author{A.E.\,Szukiel}

\affiliation{Institute for Low  Temperature and Structure Research, Polish Academy of Sciences\\
   P.O.Box 1410, 50--950 Wroclaw 2, Poland
  }

\date{\today}

\begin{abstract}
Contribution to thermal conductivity from conduction electron scattering on crystal field magnetic excitations is calculated and analyzed for normal rare-earth  inter-metallic paramagnets.
It is shown that in temperatures much lower than Debye temperature $\TD$ its behavior essentially depends on the ground state of magnetic ion in crystal field and on the excitation energy in relation to $\TD$\,.
Combined effect from the electron scattering on the crystal-field excitations, on acoustic phonons, and on nonmagnetic impurities is discussed in reference to CF splitting character and  to the
relative intensities of magnetic and non-magnetic scattering.
Total thermal conductivity resulting from these three sources of scattering is calculated for $RE$In$_3$ ($RE=\,$Nd,\,Pr,\,Tm) and compared with experimental data.


\end{abstract}
\maketitle

\section{Introduction}
Crystalline field (CF) splitting of rare earth ($RE$) ions in metals influences conduction electron transport through their interactions with 4f electrons.
In normal rare-earth systems (the ones with well localized f-level and a~stable magnetic moment) Coulomb direct and exchange interactions may be regarded
as the most important.
The last one gives magnetic contribution to transport coefficients which was studied theoretically already a~long time ago.
In early works concerning $RE$ ions as impurities in metallic systems, it was considered together with the contribution from isotropic Coulomb scattering.\cite{ful1,hi}
The formula for electrical resistivity, derived in Ref.\,[\onlinecite{hi}] in frame of Boltzmann equation solution---and that for thermal conductivity derived in Ref.\,[\onlinecite{ful1}] in  frame of the Kubo linear response theory---were then adapted for concentrated paramagnetic systems, respectively in Refs\,[\onlinecite{hes1,hes2,wong}]).
The isotropic Coulomb scattering was omitted there, as being accounted in Bloch character of conduction electrons in a~periodic lattice, and in Refs\,[\onlinecite{hes2,wong}] has been replaced by aspherical Coulomb scattering (electron-quadrupole scattering).
The magnetic contributions to the electrical resistivity,\cite{hi,hes1,hes2} and to the thermal conductivity\cite{wong} have been expressed
by a~sum of components from conduction electron scattering on the 4f-electron magnetic excitations in crystalline field (inelastic -- relating to transitions between the levels of different energies and elastic ones -- between levels of the same energy).

Calculation of the magnetic contribution to the electrical resistivity based on the formula  $\rhomag(T)$, Ref.\,[\onlinecite{hes1}] was successfully used in the interpretation of the experiment for a~number of $RE$ paramagnets, see the examples in Refs\,[\onlinecite{fou,crow,guert}].
What seems more important, some relations between the resistivity behavior and the character of CF-splitting have been established and experimentally confirmed\cite{hi,hes1,hir}.
It was found that $\rhomag(T)$ increases with temperature and in the range of temperatures corresponding to the energy of the first excited CF-level, the rate of increase depends on the excitation energy.
In high temperatures the resistivity saturation value $\rho^{\mrm{spd}}$\,, is independent of CF-splitting, while the zero temperature limit is governed by the 4f-electron ground state in the crystal field.
Through these findings, the electrical resistivity behavior may serve as a~kind of an identifier of the CF-splitting character.
The situation is different in the case of the thermal conductivity studies.
Despite it was found for some $RE$ metallic compounds that the thermal conductivity experimental results may be approximated by its electronic part\cite{wong,bau}, the role of magnetic contribution in  the electronic part behavior, as well as the behavior of magnetic contribution itself, is not clear yet.
The formula of Ref.\,[\onlinecite{wong}] was applied to the low-temperature experiment interpretation for paramagnetic (or remaining paramagnetic down to a~few Kelvins) compounds as TmSb, ErSb, PrPt$_5$, PrCu$_5$ in Ref.\,[\onlinecite{wong}] for PrCu$_5$ in Ref.\,[\onlinecite{matz}] or for PrAl$_3$ in Ref.\,[\onlinecite{mul}].
All the considered compounds exhibit similar type of CF splitting -- nonmagnetic ground state and comparable values of the first excited level energy.
Their thermal conductivity behaves also in like manner so it is understandable that there had not been any discussion about the influence of CF-splitting character on the thermal conductivity temperature dependence.
This topic was also not approached in Ref.\,[\onlinecite{ras1}], where calculations of the magnetic contribution to the thermal conductivity have been performed merely 
for two-level CF ferromagnetic system.
However, the problem seems to be the current and relevant in view of the thermal conductivity experiment for $RE$Al$_2$\cite{bau} or $RE$In$_3$\cite{much1}, among which  TmAl$_2$, PrIn$_3$, NdIn$_3$\,, or TmIn$_3$ are paramagnetic or remain paramagnetic down to very low temperature.
Their magnetic properties received attention in a~number of works, including characteristics of their crystal field.
The crystal field effects on the electrical resistivity of those compounds were also examined.
Until now, however, there are no studies analyzing thoroughly their thermal conductivity behavior and the crystal field influence on it.

We analyze this problem in the present work for  $RE$ inter-metallic paramagnets ($RE$--I--P). Their thermal conductivity we calculate  as the effect of, additively treated, contributions from the conduction electron scattering on CF-excitations, acoustic phonons and on the nonmagnetic impurities.
 Within Kohler variational approach,\cite{koh,koh1} we prove  that, for the total thermal conductivity, and for their contributions from considered types of  scattering, the  simple formula can be used -- the simpler one than that applied in Ref.\,[\onlinecite{ras1}].
For the electron--phonon contribution, we use the form after Kohler\cite{koh1} and Ziman\cite{zim}.
For the impurity contribution, the variational formula---as we show ---takes the standard form\cite{zim}, derived within the relaxation time solution of the transport equation.
The existence of relaxation time depending only on the electron energy we  also justify for the electron scattering on the CF excitations.

In consequence, the formula for the magnetic contribution to the thermal conductivity, derived by us within the variational approach proves as simple as that for the electrical resistivity derived within the relaxation time solution of transport equation in\cite{hi}.
It has simplified our analysis of the magnetic conductivity and allowed us to derive some rules  governing the conductivity behavior in low temperatures as depending on the CF-splitting character.
Combining this with the low temperature behavior of the electron--phonon and the electron--impurity contribution, we reach some general conclusions for qualitative behavior of the total thermal conductivity in this range of temperatures.
Then, we verify these conclusions by comparison  calculations  with experiment for $RE$In$_3$ ($RE=\,$Pr,\,Nd,\,Tm)\cite{much1}, and we discuss applicability of our model.
Finally, we refer critically to some aspect of calculation presented in Ref.\,[\onlinecite{ras1}], made within the similar physical model for the electron scattering and variational approach to the thermal conductivity computation.

\section[The Model]{The model and method for calculation\\ \hspace*{1.8em} of thermal conductivity}

We consider below the contributions
$\kappa^{\alpha}(T)$ to the thermal conductivity $\kappa(T)$ and
$\rho^{\alpha}(T)$ to the electrical resistivity $\rho(T)$ from conduction electron scattering on acoustic phonons ($\alpha=\mrm{ph}$), on nonmagnetic impurities
($\alpha = \mrm{imp}$) and on crystal field excitations ($\alpha = \mrm{mag}$).
Our considerations are based on the simple physical model, described in Appendices A--C.

For the conductivity and the resistivity calculation we use Kohler's formulas derived within variational method of Boltzmann equation solution\cite{koh,koh1,zim}.
We show (Appendix~A) that for the applied physical model the simplest form of these formulas, (\ref{46}) and (\ref{47})) can be used.

Applying in (\ref{46})--(\ref{47}) denotations
$\ovl{P^{\alpha}_{22}}=3\,P^{\alpha}_{22}/(P_0^{\alpha}(\kBT)^{2}\,\pi^2)$,
$\ovl{P^{\alpha}_{11}}=P^{\alpha}_{11}/P_0^{\alpha}$,
and taking into account the form of scattering matrix elements
$P^{\alpha}_{ii}(t)$, $i=1,2$, $\alpha=\mrm{ph,mag}$
(\ref{66}), (\ref{27}) depending on the reduced temperature $t=T/\TD$\,, we can represent the formulas in the form
\ba\label{7a} 
\kappa^{\alpha} &=& \kappa_0^{\alpha}\frac{t}{\ovl{P^{\alpha}_{22}(t)}}\;,
\qquad\qquad
\rho^{\alpha} = \rho^{\alpha}_0\ovl{P^{\alpha}_{11}(t)}\;,                      \nonumber\\
\kappa_0^{\alpha} &=& \frac{L_0\,\TD}{\rho_0^{\alpha}}\;,
\qquad\qquad\qquad
\rho_0^{\alpha} = \frac{P^{\alpha}_0}{J_1^2}\;,
\ea
where $L_0 = \pi^2\,\kB^2 / 3e^2$ is the Lorentz number.

The scattering matrix elements
$P^{\alpha}_{ij}$, $i,j = 1,2$ for $\alpha = \mrm{imp,mag}$ \ we derive in Appendix~B,
and for $\alpha = \mrm{ph}$ we use the form (\ref{66}).

From the Matthiessen rule
$C(\bf{k},\bf{k'}) = \sum_{\alpha} \mathit{C}^{\alpha}(\bf{k},\bf{k'})$
for the scattering probability (\ref{4}) the corresponding rule follows for the scattering matrix elements
$\ovl{P^{\alpha}_{ij}}$
and, consequently, for the total electrical resistivity
$\rho = \sum_{\alpha}\rho^{\alpha}$ and for the total thermal resistivity $W(t)=1/\kappa(t)$.
For the reduced thermal resistivity $\ovl{W(t)} = W(t)/W_0$ it takes the form
\ba\label{24} 
\ovl{W(t)} &=& \ovl{\Wph(t)} + \ovl{\Wmag(t)} + \ovl{\Wimp}\;,          \nonumber\\
\ovl{W^{\alpha}(t)} &=& W^{\alpha}(t)/W_0\;,                         \nonumber\\
W^{\alpha}(t) &=& 1/\kappa^{\alpha}(t)\;,
\ea
where  $W_0=1/\kappa^{\mrm{ph}}_0$,  according to (\ref{7a}).

Next, after substituting
$\ovl{\Pimp_{22}}$ (\ref{104}) and
$\ovl{\Pmag_{22}(t)}$ (\ref{27}) in (\ref{7a})
we get from (\ref{7a})--(\ref{24})
\ba\label{24a} 
\ovl{\Wph} =\frac{\ovl{\Pph_{22}}}{t}\;, \qquad
\ovl{\Wmag} = \cmag \frac{\ovl{\Pmag_{11}}}{t}\;, \qquad
\ovl{\Wimp} = \frac{\cimp}{t}\;, \nonumber\\
\cmag = \frac{\rhomag_0}{\rhoph_0} = \frac{\rhospd}{\rhoph_0 J(J+1)}\;, \qquad \quad
\cimp = \frac{\rhores}{\rhoph_0}\;,
\ea
where $\ovl{\Pph_{22}}(t)$ is described by (\ref{66}) and  $\rhores = \rhoimp_0 \ovl{\Pimp_{11}}$ denotes the temperature-independent impurity part of electrical resistivity (residual resistivity) and  $\rhospd$ is the high temperature saturation value of $\rhomag(t)$.

The common way to find 
coefficients $\cmag$ and $\cimp$ (\ref{24a}) for particular $RE$--I--P  relies on the assumption that the phonon contribution $\rhoph(t)$ to the resistivity of this compound  $\rho^{\mrm{M}}(t)$ and to the resistivity of its nonmagnetic counterpart, according to (\ref{7a}), are the same.
With the use of approximations
$\rhoph(t_0) = 0$ ($t_0\ll 1$),  $\rhoph(t_{\infty}) = \rhoph_0\,t_{\infty}$,  ($t_{\infty} \gg 1$)
based on, correspondingly, the low and the high temperature asymptote of $\rhoph(t)$ (\ref{7a}), one can find $\rhoph_0$ from
\ba\label{18a} 
\rho_{\sss{NM}}(t_{\infty}) &=& \rhoph_0\,t_{\infty} + \rhores_{\sss{NM}} \nonumber\\
\rho_{\sss{NM}}(t _0)       &=& \rhores_{\sss{NM}}\;,
\ea
where $\rhores_{\sss{NM}}$ denotes residual resistivity $\rhoimp$ (\ref{7a}) for the nonmagnetic compound.
For a~magnetic compound, we use the low and the high temperature asymptotics of $\rhomag(t)$ (\ref{7a})
%
\ba\label{18b} 
\rhomag(t_0) &=& (\rhomag_0/l_1) \sum_{{n=1}\atop{\vep_n=\vep_1}} a_{1n}\;,    \nonumber\\  
\rhomag(t_{\infty}) \equiv \rhospd &=& \rhomag_0\,J(J+1) .
\ea

Then, the residual resistivity for the magnetic compound $\rhores$ and $\rhomag_0$ can be found from the  equations
\ba\label{18} 
\rho(t_{\infty}) &=& \rhoph_0 \, t_{\infty} + \rhores + \rhomag_0 J(J+1)\,,                                \nonumber\\
\rho(t_{0})      &=& \rhores + \rhomag_0 \, \frac1{l_1} \sum_{{n=1}\atop{\vep_n=\vep_1}} a_{1n}\,.
\ea


\begin{table}
\caption[Tab.I]{Parameters characterizing relative impurity and magnetic contributions to the thermal resistivity.}
\begin{tabular}{|c|c|c|c|c|c|c|c|c|c|}
\hline
Compound & $J$ & $\sum_{{n=1}\atop{\vep_n=\vep_1}} a_{1n}/l_1$ & $\rhores$\,[$\mu\Omega$\,cm] & $\rhoph_0$\,[$\mu\Omega$\,cm]
& $\rhospd$\,[$\mu\Omega$\,cm] & Ref. & $\cimp$ & $\cmag$ \\
\hline\hline
$PrIn_3$   & 4   & 0   & 0.31 & 7.5 & 3.5 & ~\cite{klet3}     & 0.041 & 0.022 \\
\hline
$NdIn_3$   & 9/2 & 1.2 & 0.26 &  8  &  3  & ~\cite{hir}       & 0.03  & 0.015  \\
\hline
$TmIn_3$   & 6   &  0  & 1.8  & 7.5 & 3.1 & ~\cite{de,klet3}  & 0.24  & 0.01 \\
\hline
\end{tabular}\end{table}

Results of our calculation $\rhoph_0,\rhospd,\rhores$ and, next, $\cmag,\cimp$ with the use of (\ref{18a})--(\ref{18}) for $RE$In$_3$ ($RE=\,$Pr,\,Nd,\,Tm) are included in Table~I. \
All the calculations were done for the value $\TD=170$\,K -- following from specific heat measurements for LaIn$_3$\,\cite{diep}.
To calculate
$\sum_{{n=1}\atop{\vep_n=\vep_1}} a_{1n}/l_1$
we have used CF energies and eigenfunctions from Ref.\,[\onlinecite{lea}] corresponding to $x,W$ and $J$ included in~Table~III.
For all compounds we assumed
$t_0 = 5$\,K/170\,K, and $t_{\infty} = 300$\,K/170\,K
in correspondence to $\TD=170$\,K and to the low temperature $\rho(5$\,K) and the high temperature $\rho(300$\,K) resistivity values.
These values we have read out from the electrical resistivity data in Refs\,[\onlinecite{klet3,hir,de}]  and included in~Table~II.
For PrIn$_3$ and for NdIn$_3$ we accepted as $\rhores_{\sss{NM}}$ the values $\rho(5$\,K),
respectively for LuIn$_3$\,\cite{klet3} and for YIn$_3$\,\cite{hir}.

In the case of TmIn$_3$\,, we assumed that the compound has the same value of $\rho_0^{ph}$ as LuIn$_3$\,, adopting for the last one the value following from Ref.\,[\onlinecite{klet3}].
We based on the remark (in Ref.\,[\onlinecite{de}]) that the high temperature slopes of the experimentally obtained electrical resistivity graphs for the both compounds are the same.

In Table~II we have also included the electrical resistivity values $\rho(5$\,K), $\rho(300$\,K) obtained from the experimental graphs for $RE$In$_3$ ($RE=\,$Pr,\,Nd,\,Tm) and YIn$_3$ presented in Ref.\,[\onlinecite{much1}].
We will discuss the puzzling differences between these values of $\rho(300$\,K) and the corresponding ones for PrIn$_3$\,, TmIn$_3$\,, and YIn$_3$ in~Section~IV.

\begin{table}
\caption[Tab.II]{Comparison of the electrical resistance value for $RE$In$_3$\,, $RE=\,$Y,\,Pr,\,Nd,\,Tm, from various experimental works.
                 (*)\,These data result from our extrapolation of the graph in Ref.\,[\onlinecite{hir}] to $T=300$\,K }  \vspace{2ex}
\begin{tabular}{|c|c|c|c|c|}
\hline
Compound & $\rho(5$\,K)\,[$\mu\Omega$cm] & $\rho(300$\,K$)/\rho(5$\,K) & $\rho(300$\,K)\,[$\mu\Omega$\,cm] & Ref. \\
\hline\hline
$YIn_3$  & 0.25 & 28   & 7     & ~\cite{much1} \\
         & 0.33 & 37.6 & 13.2* & ~\cite{hir}   \\
\hline
$PrIn_3$ & 1.74 &  3.5 &  7    & ~\cite{much1} \\
         & 0.31 & 48   & 16    & ~\cite{klet3} \\
\hline
$NdIn_3$ & 1.46 & 66.3 & 17    & ~\cite{much1} \\
         & 1.5  & 51   & 16.5* & ~\cite{hir}   \\
\hline
$TmIn_3$ & 1.8  & 4.6  &  8    & ~\cite{much1} \\
         & 1.8  & 10   & 18.1  & ~\cite{de}    \\
\hline
\end{tabular}\end{table}

\section[CF Effect]{Low temperature crystal field effect\\ \hspace*{2.1em} on the thermal conductivity}
\subsection{Crystal field effect on magnetic contribution to thermal resistivity}
We focus our attention on the range of temperatures $0.03\leq t \leq 0.25$, where the thermal conductivity for $RE$In$_3$ ($RE=$\,Pr,\,Nd,\,Tm)\cite{much1}, the $RE$--I--P representatives, shows pronounced qualitative differences.
Examining the magnetic contribution to the thermal resistivity $\ovl{\Wmag}$ (\ref{24a}) in that range of temperature, we could confine our considerations to the scattering on the magnetic excitations between two lowest CF-levels.
Henceforth, we take into account only the first component of both sums in $\ovl{\Pmag_{11}}$ (\ref{27}) describing, correspondingly, the scattering on the inelastic and the elastic CF excitations.
Further, we denote by
$\ovl{a_{11}}, \ovl{a_{22}}$
the sums of weights ($a_{nm}^{zz}$) of all elastic excitations within the ground and the first excited level, and by
$\ovl{a_{12}}$  -- the sum of the weights of all inelastic excitations between these levels.
Thus, we can express $\ovl{\Wmag}$ in a~simple way, taking into account that two cases may be distinguished, depending on the $\ovl{a_{11}}$ value.

When $\ovl{a_{11}} = 0$ we deal with nonmagnetic CF ground state (NM) and denote $\ovl{\Wmag} = \ovl{\Wmag_{\mrm{NM}}}$\,.
Approximating $Z(t)\simeq 1+ \exp(-d/t)$ ($d = d_{12}$) we get
\ba\label{18c} 
\ovl{\Wmag_{\mrm{NM}}} &=& \cmag (\Win_{\mrm{NM}} + \Wel_{\mrm{NM}})\,,           \nonumber\\
   \Win_{\mrm{NM}}     &=& \frac{3\,\ovl{a_{12}}}{t\,\cosh^2[d/2t]}\,,       \nonumber\\
     \Wel_{\mrm{NM}}   &=& \frac{3\,\ovl{a_{22}}}{t\,(\exp([d/t] + 1)}\,.
\ea

$\Win_{\mrm{NM}}$ corresponds to the first sum in (\ref{27}), describing inelastic scattering, and
$\Wel_{\mrm{NM}}$ to the second one describing elastic scattering.

When $\ovl{a_{11}} \neq 0$ ($l_1>1$), the CF ground state is magnetic (M), and we denote $\ovl{\Wmag} = \ovl{\Wmag_{\mrm{M}}}$\,.
Approximating $Z(t)\simeq l_1(1+ \exp(-d/t))$ we get similarly
\ba\label{18d} 
\ovl{\Wmag_{\mrm{M}}} &=& \cmag(\Win_{\mrm{M}} + \Wel_{\mrm{M}})\,,                        \nonumber\\
    \Win_{\mrm{M}}(t) &=& \frac{3\,\ovl{a_{12}}}{l_1\,t\,\exp[d/t]\cosh^2[d/2t]}\,,       \nonumber\\
    \Wel_{\mrm{M}}(t) &=& \frac{3\,\ovl{a_{11}}}{l_1\,t(1+ \exp[-d/t])}.
\ea

\begin{table}\centering
\caption[Tab.III]{%
Characteristics of crystal field splitting for particular compounds.
$x$ and $W$ denote the CF parameters, after given references.
$\Delta$ is the energy of the excitation from the lowest level to the first excited.
$\ovl{a_{ij}}$ ($i,j = 1,2$), defined in the text, characterize the weights of elastic and inelastic excitations corresponding to the lowest and the first excited level.
For the characteristics we use CF energies and eigenfunctions presented in Ref.\,[\onlinecite{lea}], correspondingly to given $x$, $W$ and $J$. $\TD=170$\,K.}

\vspace{3ex}

\begin{tabular}{|l|ccc|ccccc|l|}
\hline
Compound        & \multicolumn{7}{|c}{CF characteristic} &  \\
\cline{2-9}     & $x$ & $W$\,[K] & {~}Ref. & Ground and first & $\ovl{a_{11}}$ \ $\ovl{a_{22}}$ & $\Delta$\,[K] & $d\!=\!\Delta/\TD$ & $\ovl{a_{12}}$ \\[-1.3ex]
   &  &  &  &  excited CF state &   &    &    &   \\
\hline\hline
PrIn$_3$  & $-0.66$  & $2.68$   &  {~~}\cite{gro}   & singlet--triplet & 0 \ \ \ 0.5  & 101 & 0.6 & 13.3\\
\hline
NdIn$_3$  & $0.31$   & $ 0.91$  & {~~}\cite{leth}   & doublet--quartet & 6.7\ \ 21.85 & 6.9 & 0.038 &5.29\\
\hline
TmIn$_3$  & $-0.665$ & $-0.57$  & {~~}\cite{mu}     & singlet--triplet & 0\ \ \ \ 0.5 & 12.9 & 0.072 & 27.97 \\
\hline
\end{tabular}\end{table}

Values of $\ovl{a_{ij}}$ ($i,j=1,2$) can be calculated with the use of 4f-electron eigenfunctions in the crystal field, presented in Ref.\,[\onlinecite{lea}].
The ones calculated for considered here $RE$In$_3$  we include in~Table~III.
As follows from the Table, the case of nonmagnetic CF ground state $\ovl{\Wmag} = \ovl{\Wmag_{\mrm{NM}}}$ involves PrIn$_3$ and TmIn$_3$, while the case of magnetic CF ground state, $\ovl{\Wmag} = \ovl{\Wmag_{\mrm{M}}}$, concerns NdIn$_3$.

We have analyzed  (\ref{18c})-(\ref{18d}) for the considered range of temperature $0.03\leq t \leq 0.25$, and the range of $d$ including values contained in~Table~III: $0.038\leq d \leq 0.6$.
We present the behavior of these functions (in said range of temperature), in dependence of $d$, in a~schematic way in~Table~IV.
Further conclusions follows from the relations
\be\label{18e} 
\frac{\Win_{\mrm{NM}}}{\Wel_{\mrm{NM}}}> 2\,\frac{\ovl{a_{12}}}{\ovl{a_{22}}}\;, \qquad \qquad
\frac{\Wel_{\mrm{M}}}{\Win_{\mrm{M}}}> \frac{\exp[d/t]}{2}\frac{\ovl{a_{11}}}{\ovl{a_{12}}},
\ee
fulfilled for the range $0.16 \leq d/t \leq 20$ corresponding to the considered values of $t$ and $d$.

As it is seen from the Table~III, in the case $\ovl{\Wmag} = \ovl{\Wmag_{\mrm{NM}}}$ (\ref{18c}), there is $\ovl{a_{12}} \gg \ovl{a_{22}}$.
It is the general property of the nonmagnetic CF ground state case, what may be verified with the use of 4f-electron eigenfunctions given in Ref.\,\onlinecite{lea}].
Hence $ \ovl{\Wmag_{\mrm{NM}}}$ may be approximated by its inelastic part  $\ovl{\Wmag_{\mrm{NM}}} \simeq \cmag\Win_{\mrm{NM}}$, (\ref{18c}), which is an increasing function for $d\geq 0.3$.
For the energy $d$ sufficiently small---with respect to lowest considered temperatures $t\geq 0.03$---the inelastic scattering becomes quasi-elastic and its contribution may be approximated by decreasing function
$\ovl{\Wmag_{\mrm{NM}}} \simeq 3 \cmag\,\ovl{a_{12}}/t$.
We have found this approximation being justified for $d\leq0.05$.

In the case $\ovl{\Wmag} = \ovl{\Wmag_{\mrm{M}}}$ (\ref{18d}) the weights of elastic $\ovl{a_{11}}$ and inelastic $\ovl{a_{12}}$ excitations may be comparable, as it is seen from the data for NdIn$_3$.
For $d$ sufficiently small, both parts $\Wel_{\mrm{M}}$ and $\Win_{\mrm{M}}$ could give a~comparable contribution to $\ovl{\Wmag_{\mrm{M}}}$, which in this case may be approximated as $\ovl{\Wmag_{\mrm{M}}} =3\cmag(\ovl{a_{12}}+ \ovl{a_{22}}/2)/(l_1\,t)$.
The component proportional to $\ovl{a_{12}}$ describes the quasi-elastic scattering, whereas the one proportional to $\ovl{a_{22}}$ describes the elastic scattering.

For the magnetic ground state and large $d$ ($d\geq 0.3$) the elastic scattering predominates, as it is seen from (\ref{18e}), and we get from (\ref{18d}) $\ovl{\Wmag_{\mrm{M}}} =3\cmag\ovl{a_{11}}/(l_1\,t)$.

The conclusion is that for large $d$ the predominant scattering is inelastic in the case of nonmagnetic ground state (NM) and is elastic in the magnetic one (M).
For small $d$ ($d\leq0.05$) the predominant scattering is quasi-elastic in the NM case or elastic and quasi-elastic in the M case.
PrIn$_3$ represents the first case, while TmIn$_3$ and NdIn$_3$ the third and the fourth cases, correspondingly.
(For TmIn$_3$ it concerns  $t\geq 0.05$,  due to $d$ value, see Table~III.)

Noting that the temperature dependence of $\ovl{\Wmag}$ for the three last cases (elastic, quasi-elastic, elastic plus quasi-elastic) differ only by coefficients,  we identify all them and describe, by contract, as \emph{elastic} scattering.
We denote for these cases $\ovl{\Wmag} \equiv \ovl{\Wmag_{\mrm{el}}}$.

Using in $\Win_{\mrm{NM}}$ (\ref{18c}) the approximation $cosh(d/2t)^2 \simeq exp(d/t)$ justified for large $d$, one can note that  $\ovl{\Wmag_{\mrm{inel}}}$ is about $\exp(d/t)$ times smaller than $\ovl{\Wmag_{\mrm{el}}} \sim 1/t$, accurate to the corresponding excitations weights ratio.  

Identifying $d \geq 0.3$ as a~large energy, and $d \leq 0.05$ as a~small one, we can summarize:

(i)~~If the electron scattering on CF excitations is predominantly inelastic (nonmagnetic $RE$ ion ground state and large excitation energy $d$) the magnetic contribution to the thermal resistivity
$\ovl{\Wmag} \equiv \ovl{\Wmag_{\mrm{inel}}}$
has the form
$\cmag\Win_{\mrm{NM}}$ (\ref{18c}),
increasingly depending on temperature.

(ii)~~If the CF-scattering is predominantly elastic (magnetic ground state and large excitation energy $d$) or quasi-elastic (nonmagnetic ground state and small $d$) or elastic plus quasi-elastic  (magnetic ground state and small $d$) the magnetic contribution to the thermal resistivity $\ovl{\Wmag }\equiv \ovl{\Wmag_{\mrm{el}}}$ is decreasing function.
For all these cases it can be approximated
$\ovl{\Wmag_{\mrm{el}}}\simeq a/t$
where $a= 3\cmag\ovl{a_{11}}/l_1 $ in the first case,
$a=3\cmag\ovl{a_{12}}$ in the second case,
and $a=3\cmag(\ovl{a_{12}}+ \ovl{a_{22}}/2)/l_1$ in the third one.

(iii)~~$\ovl{\Wmag_{\mrm{inel}}} \sim \exp[-d/t]\,\ovl{\Wmag_{\mrm{el}}}$.  


In Fig.\,1--Fig.\,3 we show  $\ovl{\Wmag}$\, (\ref{24a}) calculated for all crystal field levels corresponding to CF splitting for PrIn$_3$\,, NdIn$_3$\,, and TmIn$_3$\,, (according to $x$, $W$ given  in Table~III).
As $\cmag$ we use the values included in Table~I.
The graphs $\ovl{\Wmag}$ confirm (i)-(iii), however for TmIn$_3$ (Fig.\,3) it concerns  $t\geq 0.05$.
Only for this range of temperature the CF-scattering for this compound is quasi-elastic and TmIn$_3$ may be included into the same group (ii), to which also NdIn$_3$ belongs.
For both compounds $\ovl{\Wmag }\equiv \ovl{\Wmag_{\mrm{el}}}$  is approximated by  the  same formula
$\ovl{\Wmag}\simeq \ovl{\Wmag_{\mrm{el}}} =a/t$,
where $a=3\cmag\ovl{a_{12}}$ corresponds to TmIn$_3$,
whereas  $a=3\cmag(\ovl{a_{12}}+ \ovl{a_{22}}/2)/l_1$ corresponds to NdIn$_3$.
Calculating $a$ for both compounds, with the use of data from Table~I and Table~III,  we get the greater value for TmIn$_3$\,.
The same relation holds for the values of $\ovl{\Wmag}$ for these compounds, what is seen comparing corresponding graphs in Fig.\,2 and Fig.\,3.
We can recognize this as an evidence of the correctness of our low-temperature approximation $\ovl{\Wmag_{\mrm{el}}}\simeq a/t$ in case (ii), at least with respect to TmIn$_3$ and NdIn$_3$\,.

\begin{table}
\centering
\caption[Tab.IV]{%
Type of monotonicity for inelastic and elastic parts of magnetic contribution to the thermal resistivity $\ovl{\Wmag(t)}$ (\ref{24a}) approximated for $0.03 \leq t \leq 0.25$ for magnetic ($W_{\mrm{M}}$) and nonmagnetic ($W_{\mrm{NM}}$) ground CF state.
Symbols $\nearrow$, $\searrow$ denote respectively the function increasing or decreasing in the whole interval.
Description \emph{max} denotes maximum of the function within the interval.}
\vspace{3ex}
\begin{tabular}{|c|c|c|c|c|l}
\hline
 & \qquad\ $\Win_{\mrm{NM}}$ \qquad\ & \qquad\ $\Wel_{\mrm{NM}}$ \qquad\ & \qquad\ $\Win_{\mrm{M}} $ \qquad\ & \qquad\ $\Wel_{\mrm{M}}$ \qquad\ \\
\hline\hline
$d\geq0.3$ & $\nearrow$& $ \nearrow$ & $\nearrow$ & $\searrow$ \\
\hline
  $0.05 < d < 0.3$ \qquad & max & max & max & $\searrow$ \\
\hline
$d \leq 0.05$ & $\searrow$ & $\searrow$ & $\searrow$ & $\searrow$ \\
\hline
\end{tabular}
\end{table}

\subsection{Crystal field effect on total thermal conductivity}
The low temperature behavior of the thermal conductivity for considered here paramagnetic $RE$In$_3$ ($RE =\,$Pr,\,Tm,\,Nd)\cite{much1}, is similar to the behavior observed in their nonmagnetic counterparts YIn$_3$ and LuIn$_3$\,.
The conductivity of YIn$_3$ shows maximum\cite{much1} like PrIn$_3$ and NdIn$_3$, while that of LuIn$_3$ increases monotonously\,\cite{much2}.  

For nonmagnetic metals, both types of behavior have long been known and were explained in Ref.\,[\onlinecite{wil}] within the employed here model.
Thus, we can repeat the arguments presented there with the use of $\ovl{\kappa(t)}=1/\ovl{W(t)}$, (\ref{24}), after omitting magnetic scattering.
Approximating $\ovl{\Wph(t)} $ (\ref{24a}) by its low-temperature form
$\ovl{\Wph(t\to0)}$ we get
\ba\label{50}
\ovl{\kappa(t)} &=& \left[ \ovl{\Wph} + \ovl{\Wimp} \right]^{-1},                   \nonumber\\
\ovl{\Wph}   &\simeq& 12\,\cJ_5(\infty) \, n_s \; \frac{t^2}{\pi^2}\;,
\ea
where $\cJ_5(\infty) \simeq 125$. ($\cJ_5(t)$ and $n_s$ are defined in (\ref{66})).
Behavior of $\ovl{\kappa(t)}$ (\ref{50}) is the effect of a~competition between inelastic scattering on phonons -- described by increasing function $\ovl{\Wph}$ (\ref{50}) and elastic scattering on impurities -- described by decreasing function $\ovl{\Wimp}$ (\ref{24a}).

It can be easily verified that $\ovl{\kappa(t)}$ (\ref{50}) exhibits maximum at
\be\label{52}
t_{\max} = \left( \frac{\pi^2\,\cimp}{3\,000\;n_s} \right)^{1/3}.
\ee

Requirement that the maximum occurs inside the considered interval of temperatures,
$0.03 \leq t_{\max} \leq 0.25$,
gives the upper and the lower limit for parameter $\cimp / n_s$\,.
For a~fixed  $n_s$\,, there is a~threshold value for $\cimp = \rhores / \rhoph_0$, beyond which the maximum disappears in this range of temperatures, and $\ovl{\kappa(t)}$ becomes an increasing function.
For fixed $\rhoph_0$, we can consider the threshold value for $\rhores$.

A~direct conclusion from (\ref{50}) is that $\ovl{\kappa(t)}$ adopts greater values for smaller $\cimp$ or smaller~$n_s$\,.
Applying numerical analysis, we have found that the maximum becomes more pronounced for smaller $\cimp$, but flattens for smaller $n_s$.

Consider now the low temperature behavior of the thermal conductivity for magnetic compounds -- described by the formula
$\ovl{\kappa(t)} = 1/\ovl{W(t)}$ (\ref{24}).
Recalling (i)--(ii) from the previous subsection, we can note that $\ovl{\Wmag_{\mrm{inel}}}$ behaves similarly to $\ovl{W^{ph}(t)}$ (\ref{50}) and
$\ovl{\Wmag_{\mrm{el}}}$  behaves in the same way as $\ovl{\Wimp}$.
%
%
It is also seen, in Fig.\,1--Fig.\,3 for $\ovl{\Wmag} \equiv \ovl{\Wmag_{\mrm{inel}}}$ corresponding to PrIn$_3$  and for $\ovl{\Wmag} \equiv \ovl{\Wmag_{\mrm{el}}}$ corresponding to NdIn$_3$\,, TmIn$_3$\,,  where graph of $\ovl{W^{ph}(t)}$ represents the non-approximated form of the phonon contribution (\ref{24a}).
Combining above comments with the theory for nonmagnetic metals\cite{wil}, we get its extension for magnetic metals ($RE$--I--P).

\begin{figure}\begin{center}
\includegraphics[angle=270,width=0.65\columnwidth]{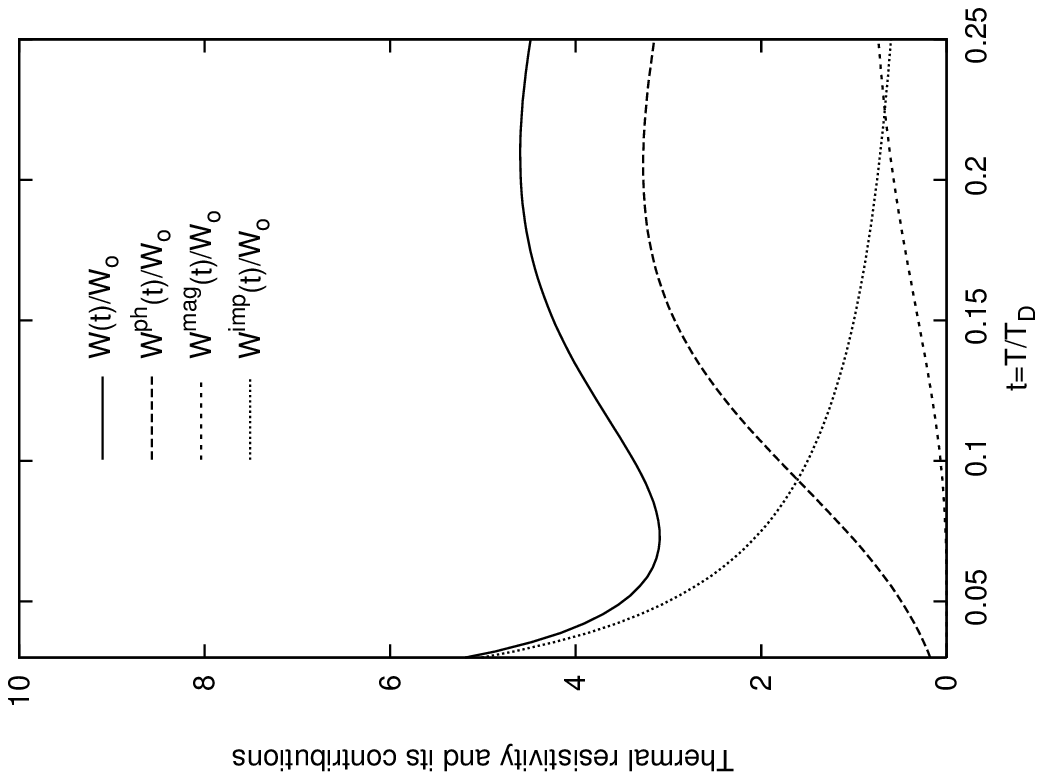}
\caption[Fig.1]{%
Reduced thermal resistivity $W(t)/W_0 \equiv \ovl{W(t)}$ (\ref{24}) and its contributions $W^{\alpha}(t) / W_0\equiv \ovl{W^{\alpha}(t)}$ (\ref{24a}) for PrIn$_3$. \ $n_s=1.3$,\, $c_{mag}=0.022$,\  $c_{imp}=0.15$. }
\end{center}\end{figure}

\begin{figure}\begin{center}
\includegraphics[angle=270,width=0.65\columnwidth]{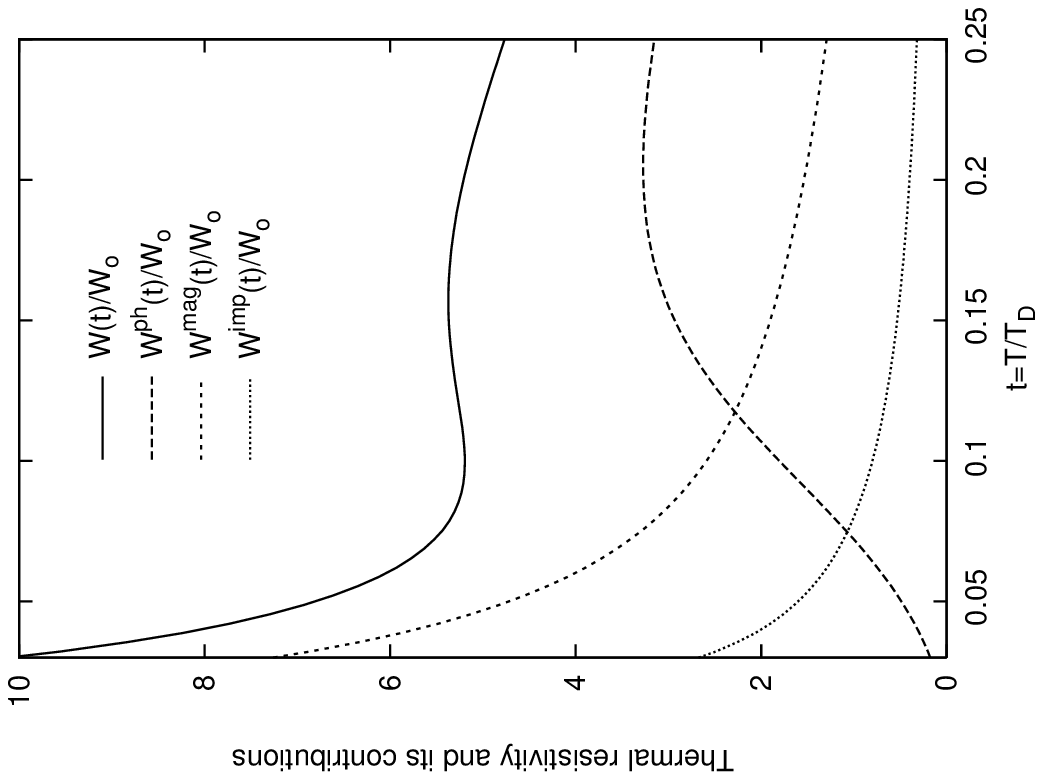}
\caption[Fig.2]{%
Reduced thermal resistivity  $W(t)/W_0\equiv \ovl{W(t)}$ (\ref{24}) and its contributions $W^{\alpha}(t)/W_0\equiv \ovl{W^{\alpha}(t)}$ (\ref{24a})  for NdIn$_3$. $n_s=1.3$,\, $c_{mag}=0.015$,\,$c_{imp}=0.08$.}
\end{center}
\end{figure}

\begin{figure}
\begin{center}
\includegraphics[angle=270,width=0.65\columnwidth]{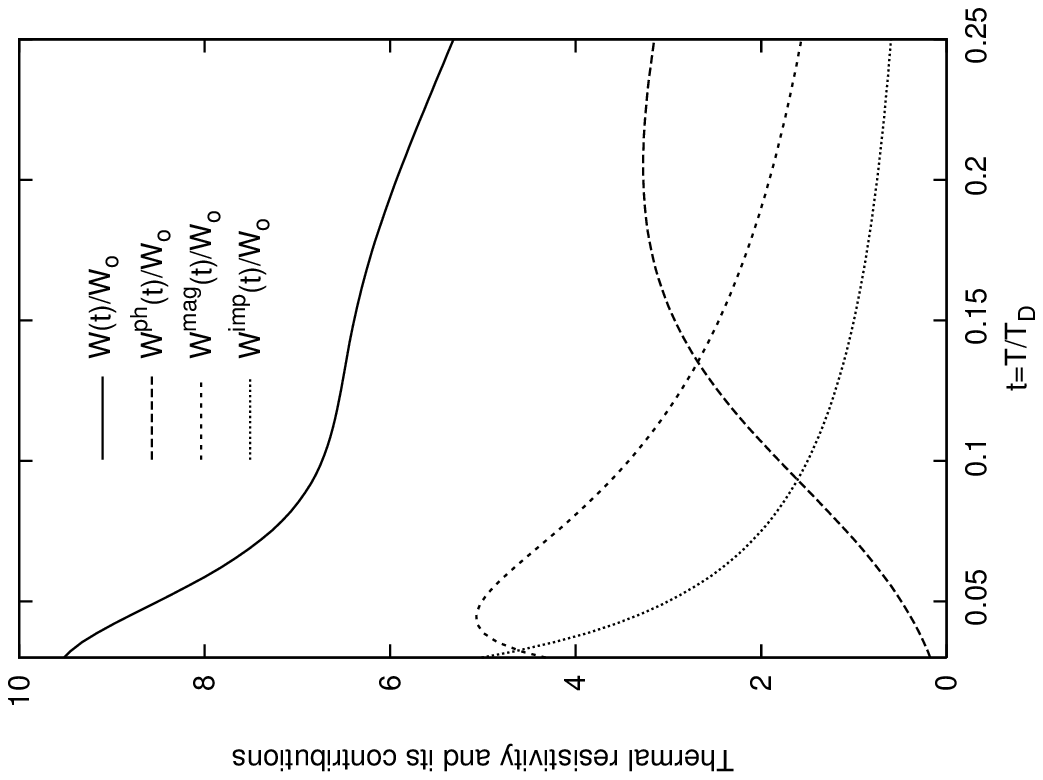}
\caption[Fig.3]{%
Reduced thermal resistivity $W(t)/W_0\equiv \ovl{W(t)}$ (\ref{24}) and its contributions $W^{\alpha}(t)/W_0\equiv \ovl{W^{\alpha}(t)}$ (\ref{24a}) for TmIn$_3$. $n_s = 1.3$,\, $\cmag = 0.01$,\,$\cimp = 0.15$.}
\end{center}\end{figure}

(I)~In systems with a~predominance of inelastic CF scattering (see (i) in Section III\,A), the thermal conductivity exhibits the low temperature maximum when the impurity contribution to the scattering $\ovl{\Wimp}$\, is sufficiently small with respect to the contribution from the sum of $\ovl{\Wph}$ and $\ovl{\Wmag_{\mrm{inel}}}$.

(II)~In systems with a~predominance of elastic or quasi-elastic (or elastic plus quasi-elastic) CF scattering (see (ii) in Section III\,A),
the thermal conductivity exhibits maximum when the sum of contributions from  $\ovl{\Wimp}$ and  $\ovl{\Wmag_{\mrm{el}}}$ is sufficiently small with respect to $\ovl{\Wph}$.

Realizing that $\ovl{\Wmag_{\mrm{inel}}}$  acts like reducing   while  $\ovl{\Wmag_{\mrm{el}}}$  acts like  enlarging $\ovl{\Wimp}$ in nonmagnetic metals, it is easy to see that $t_{\max}$ -- the temperature of the maximum and its value depend on $\cimp$ in the similar way like for nonmagnetic metals.
Therefore, in each case~(i) and~(ii), for a~fixed $n_s$ and $\cmag$ there is a~\emph{sufficiently small} value of $\cimp$ required for the conductivity maximum in the range of temperatures $0.03 \leq t\leq 0.25$.
Similarly, there must be a~threshold value for $\cimp = \rhores / \rhoph_0$ (a~threshold value for $\rhores$ when $\rhoph_0$ is fixed), beyond which the maximum disappears in the considered range of temperatures, and the conductivity becomes an increasing function.

In contradistinction to the nonmagnetic metals case, we cannot establish its value analytically.
However, for $RE$--I--P of the same phonon contribution (fixed $n_s$ and $\rhoph_0$) some conclusions follow from (I)--(II) combined with (i)--(iii):

(R1)~In case~(I), the threshold for $\cimp$  is the larger the greater is $\ovl{\Wmag_{\mrm{inel}}}$ (the greater is $\cmag\,\ovl{a_{12}}$ and smaller is $d$).
In case (II) the threshold is the larger the smaller is $\ovl{\Wmag_{\mrm{el}}}$ (the smaller is $a$).

(R2)~The threshold in case~(II) should be smaller than that in case~(I), independently of values of $\cmag$ corresponding to $\ovl{\Wmag_{\mrm{el}}}$ and $\ovl{\Wmag_{\mrm{inel}}}$ in these cases.

Applying (R1) to NdIn$_3$ and TmIn$_3$ we can conclude that the threshold for the first compound should be greater, because $\ovl{\Wmag_{\mrm{el}}}$ is greater for TmIn$_3$\,.
It follows from the greater value of $a$ for the last compound, which we have noted in the previous subsection, and what is seen from comparison of the graphs $\ovl{\Wmag(t)}$ in Fig.\,2 and Fig.\,3.
Applying (R2) to $RE$In$_3$ ($RE=\,$Pr,\,Nd,\,Tm) we may expect the threshold for PrIn$_3$ being greater than for the other compounds.
Our numerical findings confirm these relations, although the difference between thresholds found for NdIn$_3$ and TmIn$_3$ is minute.
The obvious conclusion from relations between thresholds is that the conductivity of PrIn$_3$ may exhibit the more pronounced maximum for the value of $c_{imp}$\, ($\rhores$) greater than a~value corresponding to the conductivity maximum for NdIn$_3$\,.
On the other hand, the value $c_{imp}$ ($\rhores$) for which the conductivity of TmIn$_3$ increases may be sufficiently small for the conductivity of PrIn$_3$ to exhibit a~maximum.

The above conclusion is confirmed by the conductivity experiment for $RE$In$_3$ ($RE=\,$Pr,\,Nd,\,Tm)\cite{much1}, when one takes into account the residual resistivity values $\rhores(\mrm{NdIn}_3) = 0.26\,\mu \Omega$\,cm,
$\rhores(\mrm{PrIn}_3) = 1.74\,\mu\Omega$\,cm,
$\rhores(\mrm{TmIn}_3) = 1.8\,\mu\Omega$\,cm\, (obtained from (\ref{18}) with  $\rho(t_{0})=\rho(5$\,K) included in~Table~II, 
$\rhomag_0$ from Table~I and
$\sum_{{n=1}\atop{\vep_n=\vep_1}} a_{1n}/l_1$
corresponding to CF splitting data from Table~III.

We also illustrate this with the calculation of $\ovl{W(t)}$ (\ref{24}), which is seen in Fig.\,1--Fig.\,3, when one takes into account the equivalence between the conductivity maximum and the resistivity minimum, and between the conductivity increase and the resistivity decrease regions.

As the additional difference between the conductivity for PrIn$_3$ and for both NdIn$_3$ and TmIn$_3$ we can note its much greater values than for the other compounds.
It is seen from the experiment data presented in Fig.\,4, and also from results of our calculations of $\ovl{W(t)}$ (\ref{24}) in Fig.\,1--Fig.\,3, and is a~consequence of relation~(iii) in Section~III\,A:  $\ovl{\Wmag_{\mrm{inel}}}\ll \ovl{\Wmag_{\mrm{el}}}$.

We can generalize these characteristics of the crystal field influence on the conductivity for $RE$In$_3$ ($RE=\,$Pr,\,Nd,\,Tm) for arbitrary group of $RE$--I--P of  the same $\ovl{\Wph}$.
The general conclusion appropriate for such compounds is that the conductivity of compounds of type~(i) is more likely to exhibit a~maximum and larger values than the conductivity of the compounds of type~(ii).

\begin{figure}
\begin{center}
\includegraphics[angle=270,width=0.65\columnwidth]{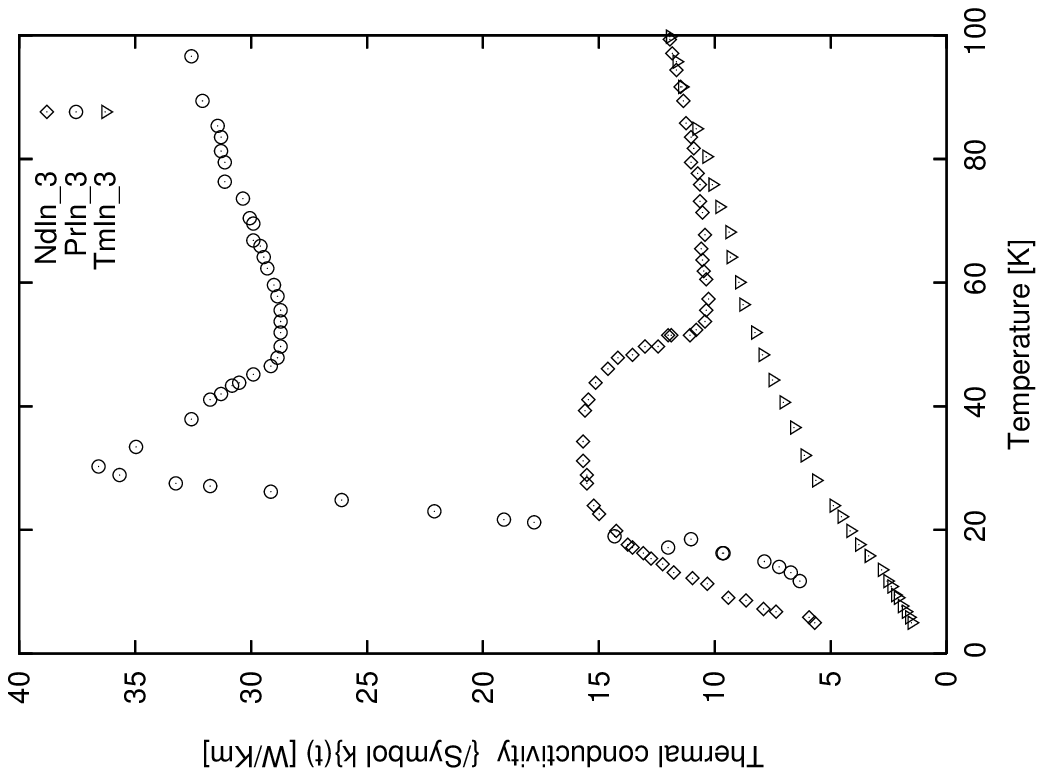}
\caption[Fig.4]{%
Thermal conductivity for PrIn$_3$\,,  NdIn$_3$\,, and TmIn$_3$ after\cite{much1}.}
\end{center}\end{figure}

\section{Comparison with experiment}
With the use of the theory presented in the previous section and the results of calculation of $\ovl{W}$ in Fig.\,1--Fig.\,3 we explain the qualitative differences between the thermal conductivity experiment for $RE$In$_3$ ($RE=\,$Pr,\,Nd,\,Tm)\cite{much1}, likewise the relations between the conductivity values.
These results, however, cannot be used for a~thorough comparison with the experiment~\cite{much1}.

\begin{figure}\centering
\includegraphics[angle=270,width=0.65\columnwidth]{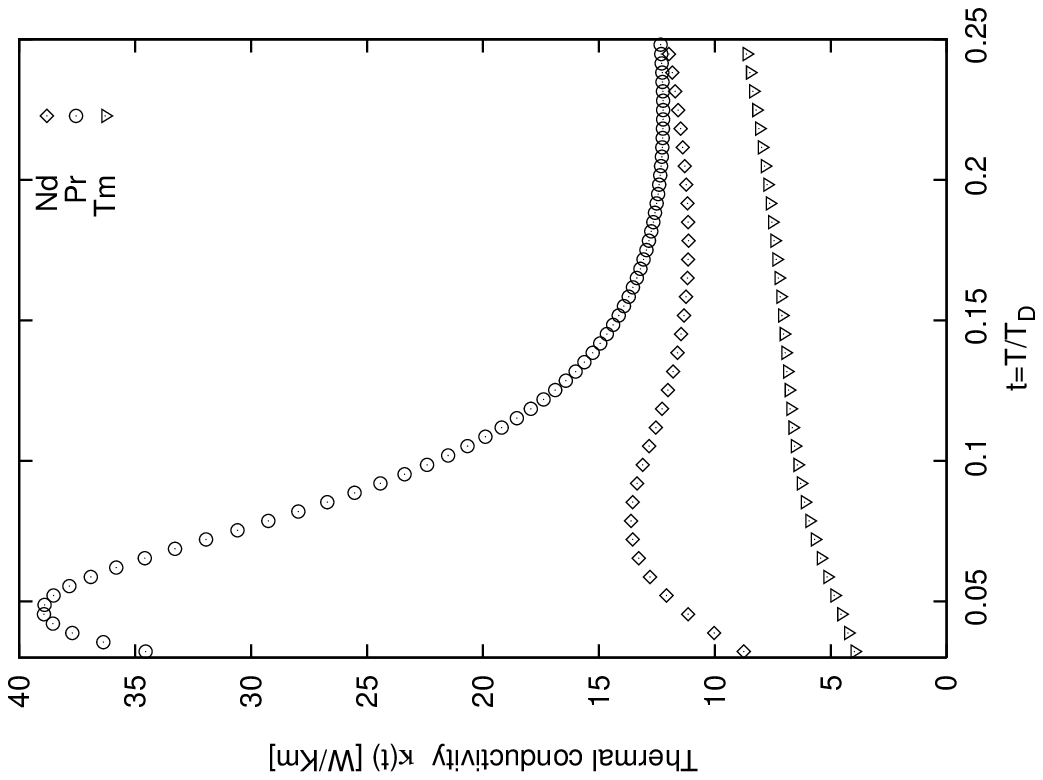}
\caption[Fig.5]{\raggedright
Thermal conductivity  $\kappa(t) = 1 / (W_0\,\ovl{W(t)})$ (\ref{24}) for CF splitting corresponding to PrIn$_3$\,, NdIn$_3$ and TmIn$_3$ described in~Table~III, $\cmag,\cimp$ from Table~I, and $n_s = 1.3$.  $\kappa_0^{\mrm{ph}} = 52.1$\,[W/mK] corresponds to $\rhoph_0 = 8\,[\mu\Omega$\,cm].}
\end{figure}

For this purpose the calculations should be done with
$\cimp = \rhores / \rhoph_0$,  $\cmag = \rhospd / \rhoph_0\, J(J+1)$
following from the electrical resistivity experiment for $RE$In$_3$\,, $RE=\,$Pr,\,Nd,\,Tm, as well as for their nonmagnetic counterpart YIn$_3$\,, which was performed with the use of the same samples as were used for the conductivity results in Ref.\,[\onlinecite{much1}], and was presented therein.
Following the approach described at the end of Section~II, we have got  $\rho(t_0)$ and $\rho(t_{\infty})$ reading out the low temperature $\rho(5$\,K) and the high temperature $\rho(300$\,K) resistivity values from the graphs\cite{much1}.
As $\rhores_{\mrm{NM}}$ we have accepted the value $\rho(5$\,K) for YIn$_3$.
(All these values are included in~Table~II).
However, substituting $\rhores_{\mrm{NM}}$ and $\rho(t_0)$, $\rho(t_{\infty})$  for PrIn$_3$\, and TmIn$_3$, to (\ref{18a})--(\ref{18}) we have ended up in contradictions.  
Therefore the method which works for the resistivity data obtained from Refs\,[\onlinecite{de,klet3}], fails for the data obtained from Ref.\,[\onlinecite{much1}].
If one regards that in the last paper $\rho(300$\,K) for PrIn$_3$\,, TmIn$_3$\,, and YIn$_3$ is about two times smaller than the respective value in Refs\,[\onlinecite{hir,de,klet3}] (see Table~III), the failure of the method is not such surprising as the results of Ref.\,[\onlinecite{much1}] themselves.

Comparing the residual-resistance ratios (RRR) $\rho(300$\,K$)/\rho(5$\,K)  and the residual-resistance values $\rhores \simeq \rho(5$\,K) included in Table~II, one can notice that, for PrIn$_3$\,, the results presented in Ref.\,[\onlinecite{much1}] are obtained with the use of a~sample of much lower purity than the one   presented in Ref.\,[\onlinecite{klet3}].
It may be an explanation of the great differences between the high temperature values of the resistivity for this compound obtained in Ref.\,[\onlinecite{much1}] and in Ref.\,[\onlinecite{klet3}].
As it was argued in Ref.\,[\onlinecite{zim}], a~large contribution of residual resistivity to the total resistivity essentially influences the character of electron--phonon scattering and makes this scattering dependent on the electron--impurity scattering.
For such a~case, our way of calculation based on the Matthiessen rule, including equations (\ref{18a})--(\ref{18})), may be inappropriate.
Such explanation, however, does not fit to the results for TmIn$_3$\,, and YIn$_3$\,.
The RRR values for these compounds\cite{much1} are about half of the values following from Refs\,[\onlinecite{klet3,hir}], whereas the values $\rhores \simeq \rho(5$\,K)\cite{much1} are comparable with the corresponding ones in the first two papers.
We must recognize the differences between the results for TmIn$_3$ and YIn$_3$ in Ref.\,[\onlinecite{much1}] and in Refs\,[\onlinecite{klet3,hir}] as unexplained.
There is a~different case with NdIn$_3$\,, for which---in contrast to the other compounds---all the electrical resistivity values in Ref.\,[\onlinecite{much1}] agree very well with those in~Ref.\,[\onlinecite{hir}].
It justified, in our opinion, accepting the values $\cimp,\cmag$ in Table~I as corresponding also to the sample of NdIn$_3$ used in~ Ref.\,[\onlinecite{much1}].
Therefore, among the results of our calculation of $\kappa(t) = 1 / (W_0\,\ovl{W(t)})$ (\ref{24}) presented in Fig.\,5, the only reliable comparison with the experiment\cite{much1} may be performed for NdIn$_3$\,.
Comparing  graph of $\kappa(t)$  for this compound in Fig.\,5 with the experimental behavior of thermal conductivity in Fig.\,4, and taking into account the temperature scaling $t = T/\TD$ with $\TD = 170$\,K one can note that:

--- $\kappa(t)$ in Fig.\,5 has the maximum at temperature about twice smaller than the experimental one $t_{\max} \simeq 0.17$, and for the temperature of the  minimum $t_{\min}\simeq 0.37$ the difference is greater.

--- values of $\kappa(t)$ are smaller than the corresponding experimental ones, and this disagreement, being about 20\,\%\ at the temperature of $\kappa(t)$ maximum, reaches 30\,\%\ for other temperatures.

It should be noted that we did not obtain better agreement changing value of $n_s$\,.  
We have found, similarly as in the case of non-magnetic compounds discussed in~Section~III\,A, that increase of $n_s$ moves the maximum towards lower temperatures, and that decrease of $n_s$ cause its flattening.

There are several possible causes of discrepancies described above.
As the crucial one, we consider deficiencies of the standard model for the electron--phonon scattering and the electron scattering on nonmagnetic impurities, which we used in our calculation.

In the simplest way it may be demonstrated by comparing $\kappa(t)$ (\ref{50}) with the experiment for YIn$_3$\,\cite{much1}.
The thermal conductivity for this compound shows the maximum at $t_{\max} = 0.17$ assuming $\TD = 170$\,K, as we have assumed in the calculations of previous section.
Substituting this in (\ref{52}) together with $\cimp \simeq 0.06$ -- following from the resistivity data in Ref.\,[\onlinecite{much1}], or together with $\cimp \simeq 0.03$ -- following from the data in Ref.\,[\onlinecite{hir}], one obtains the nonmetallic values of~$n_s$\,.
On the other hand, for metallic values ($n_s \geq 0.63$) and  for\, $0.03 \leq \cimp \leq 0.06$\, one gets $\TD > 400$\,K which seems to be unrealistic.
The general arguments for the theory oversimplification can be found in Refs\,[\onlinecite{zim,bla,ber}].
They refer to the electronic transport in non-magnetic metals, but they remain relevant also for the magnetic ones considered in the present paper.
The unclear sense of parameter $n_s$ with respect to these metals, which we have mentioned in Ref.\,[\onlinecite{uw}] is the one example.
Another one is the application of the Matthiessen rule.
We have recognized it as inappropriate for calculation of thermal conductivity in case of low purity sample of PrIn$_3$ in Ref\,[\onlinecite{much1}], but its application is an approximation in every case, and leads to a~discrepancy with the experiment.
We may expect this discrepancy increase due to the scattering on magnetic CF excitations.
As follows from the thermal conductivity calculation for the CF-split $RE$ impurities in a~metal\cite{ful1}, there are essential differences in the thermal conductivity values depending whether the Matthiessen rule is applied or not.

As the additional cause of poor agreement with the experiment in the case of NdIn$_3$ we may consider the omitting---in our calculation---the aspherical Coulomb interaction as a~source of electron scattering.
The important role of that scattering for the electrical resistivity for NdIn$_3$ has been proven in~Ref.\,[\onlinecite{hir}].

In the case of PrIn$_3$ and TmIn$_3$\,, although the results of our calculation presented in Fig.\,5 have no direct relevance to the experiment~\cite{much1}, they agree qualitatively with that experiment.
It concerns both the conductivity behavior for particular compound as well as relations between the conductivity values for all $RE$In$_3$ ($RE=\,$Pr,\,Nd,\,Tm).
We can conclude that the CF-scattering effects in these compounds are strong enough to make their conductivities less sensitive to the effects of the non-additivity of the scattering intensities.

\section{Summary and conclusions}

We have analyzed how the conduction electron scattering on magnetic crystal-field (CF) excitations influences the low temperature dependence of the thermal conductivity in rare-earth intermetallic paramagnets ($RE$--I--P).
We have considered the thermal conductivity as resulting from the conduction electrons independently scattered by acoustic phonons, by the non-magnetic impurities and by the CF magnetic excitations.\, 
Hence we have applied the 
formula
$\kappa=1/(W^{ph} + W^{imp} + W^{mag})$,
where $W^{\alpha}$  the $\alpha=ph, imp, mag$ are contributions to the thermal resistivity $W=1/\kappa$.
They are proportional to the material parameters  $\rhoph$,  $\rhores$, $\rhomag$, which can be found from the electrical resistivity experiment in a way based on   the Mathiessen rule, what we have described in Section~I.

%
Calculating the magnetic contribution and examining its behavior below $0.25\,T_{D}$\,, we have found that it increases like $W^{ph}$ when the magnetic scattering is dominated by the inelastic CF excitations and decreases like $W^{imp}$ when the elastic and/or quasi-elastic scattering prevails.
In the first case~(i) we denote $W^{mag}= W^{mag}_{inel}$, in the second one~(ii) $W^{mag}= W^{mag}_{el}$\,, considering also the quasi-elastic scattering as the \emph{elastic} one.  We have found that $W^{mag}_{inel}\ll W^{mag}_{el}$ and established dependence of  $W^{mag}_{inel}$ and $W^{mag}_{el}$ on some parameters characterizing CF splitting.
The case~(i) concerns compounds like PrIn$_3$\,, for which the f-electron ground state is nonmagnetic and $d$, the energy of the first excited CF state, is sufficiently large in comparison with $0.25\,\TD$\,.
The case~(ii) concerns compounds where $d$ is sufficiently small, like NdIn$_3$ (with magnetic ground state) or TmIn$_3$ (with nonmagnetic ground state).
We illustrated this in Fig.\,1--Fig.\,3 calculating $\overline{W^{mag}}= W^{mag}/W_0$ (\ref{24})--(\ref{24a}).

From these findings about $W^{mag}$ we could conclude that values and behavior of the total thermal conductivity in $RE$--I--P, like those in nonmagnetic metals,  results from a~competition between contributions from inelastic and elastic scattering.
The inelastic scattering is described by $W^{ph}$  and in case~(i) additionally by $W^{mag}_{inel}$, while the elastic scattering is described by $W^{imp}$ and in the case~(ii), additionally, by $W^{mag}_{el}$.
When the elastic scattering is sufficiently small with respect to the inelastic scattering, the conductivity exhibits low temperature maximum, otherwise it  increases. 
For a~particular $RE$--I--P, like for nonmagnetic metals, the temperature of the maximum increasingly depends on $\cimp$\,, or, equivalently, on the value of $\rhores$ -- both determining the impurity contribution.
However, it depends not only on $W^{ph}$, as in the case of nonmagnetic metals, but also on $W^{mag}$.
The same concerns the threshold for $c_{imp}$ ($\rhores$)\,, beyond which the maximum disappears.
The dependencies of the thresholds on parameters characterizing CF-splitting we described in~(R1) Section~III for both cases~(i) and~(ii).
We have also established (in (R2) therein) that the threshold value in case~(i) is greater than in case~(ii) and that the same relation concerns the conductivity values.
These relations, which apply to $RE$--I--P of the same phonon contribution, lead to the conclusion that the conductivity maximum and its large values require much purer samples in the case of predominately elastic and/or quasi-elastic CF-scattering than in the case of the inelastic scattering predominance.
It can be seen analyzing the conductivity experiment for $RE$In$_3$ ($RE=\,$Pr,\,Nd,\,Tm)\cite{much1}, and corresponding values of $\rhores$, as we have discussed in~Section~III.
We have also illustrated it by calculating the reduced thermal resistivity $\overline{W(t)}$ for these compounds, presented in Fig.\,1--Fig.\,3.

In these calculations we could not use the values $\cimp=\rhores/\rhoph$ corresponding to the experimental data in Ref.\,[\onlinecite{much1}], because finding a~consistent set of material parameters $\rhores$, $\rhoph$, $\rhomag$, based on the electrical resistivity results in that paper for $RE$In$_3$ ($RE=\,$Pr,\,Nd,\,Tm) and for their nonmagnetic counterpart YIn$_3$\,, proved impossible.
As we note for PrIn$_3$\,, TmIn$_3$\,, and YIn$_3$ the resistivity experimental values in Ref.\,[\onlinecite{much1}] differ greatly (in the whole investigated range of temperatures) from the corresponding ones in Refs\,[\onlinecite{hir, klet3, de}].
In case of PrIn$_3$\,, basing on the residual-resistance ratio (RRR) and $\rhores$ for the sample used in the experiment\cite{much1}, we could ascribe this difference to the Matthiessen rule breaking caused by low purity of the sample.
In such conclusion we followed the arguments of Ref.\,[\onlinecite{zim}].  For TmIn$_3$ and YIn$_3$ we had to recognize the differences as unexplained.

The consistent set of material parameters and, consequently, the coefficients $\cmag$ and $\cimp$ for $RE$In$_3$ ($RE=\,$Pr,\,Nd,\,Tm) could be found in the electrical resistivity experiment of Refs\,[\onlinecite{hir, klet3, de}].
These coefficients (included in Table~I) we used in our calculation of $\kappa(t)=1/(W_0\,\overline{W(t)})$ (\ref{24})-(\ref{24a})  and presented results in Fig.\,5.
The only reliable comparison of these results with the experiment\cite{much1} we could do for NdIn$_3$\,, for which the values of the electrical resistivity in\, Ref.\,[\onlinecite{much1}] agree with those in Ref.\,[\onlinecite{hir}].
From discussion of discrepancies between the results of our calculation in this case and the experiment\cite{much1} one can conclude that our model of calculation is too simple to provide a~good quantitative agreement.
The obvious  way of improving the model is to go beyond Mathiessen rule in calculation, what seems to be a~nontrivial problem to solve.
The other improvement may be to take into account the aspherical Coulomb scattering as the additional source of scattering, but it seems to be of specific (for NdIn$_3$), not of a general importance for $RE$--I--P.
Nevertheless, the results of calculation presented in Fig.\,5, also for PrIn$_3$ and TmIn$_3$\,---although their quantitative agreement with the experiment\cite{much1} is worse than that for NdIn$_3$---incline us to believe that the model may be useful for qualitative interpretation of thermal conductivity of any $RE$--I--P.
It proves to be particularly useful for analyzing the CF effects on the differences in the conductivity behavior for a~group of iso-structural $RE$--I--P, as we have shown in our discussion performed for $RE$In$_3$\,.

It should be stressed that all these capabilities of the model are based on the simplicity of the variational formula for the conductivity (\ref{46}), much simpler than that was used in the thermal conductivity calculation for ferromagnetic $RE$ intermetallics in Ref.\,[\onlinecite{ras1}].
As we have shown in Appendix~B, the term $-T\,S^2\,L_{EE}$, of which the latter formula differs from (\ref{46}) is negligible for the physical model we have considered.
It can be similarly demonstrated that this term is also negligible for the model considered in Ref.\,[\onlinecite{ras1}].
In that paper it was calculated with the use of incorrect form of the electron--phonon contribution to the thermoelectrical power, $S^{ph}$ obtained in Refs\,[\onlinecite{durcz1, durcz2}].
It caused, as it seems, its overestimation.
We have proved incorrectness of the formula for $S^{ph}$\,\,\cite{durcz1,durcz2} in Ref.\,[\onlinecite{asz}].

\appendix 
\section{} 
To get the formula for thermal $\kappa$ and electrical $\sigma$ conductivities it is usual to start from  linear relations between the electrical and the thermal currents on the one hand, and the electrical field and the temperature gradient on the other hand, with the coefficients $L_{\mrm{ET}}$\,, $L_{\mrm{TE}}$\,, $L_{\mrm{EE}}$\,, $L_{\mrm{TT}}$\,,( the generalized transport coefficients), see Ref.\,[\onlinecite{zim}], Chapt.\,VII.
With the use of expression $S = -L_{\mrm{ET}} / L_{\mrm{EE}}$ for the thermoelectric power one can write
\ba\label{1} 
\kappa &=&  -T\,S^2\,L_{\mrm{EE}} -L_{\mrm{TT}}\,,     \nonumber\\
\sigma &=& L_{\mrm{EE}}\;.
\ea

The transport coefficients can be found with the use of the Kohler's variational method for the Boltzmann equation\cite{koh,koh1,zim}.
Its solution, the non-equilibrium distribution function, is represented there as the linear combination of some basis $\phi_i(\bf{k})$, $i=1,\ldots,n$ and the coefficients of this combination are found from the so called variational principle.
After decomposition  the distribution function with respect to the basis one gets the generalized transport coefficients expressed by trial currents  $J_i , U_i$
\ba\label{2} 
J_i &=& - e \int\!\di\bk \left(\frac{-\partial f^0_k}{\partial\vep_k} \right) \phi_i(\bk) (\bv \cdot \bu)\;,      \nonumber\\
U_i &=& -e \int\!\di\bk \left(\frac{-\partial f^0_k}{\partial\vep_k} \right) \phi_i(\bk)(\bv \cdot \bu)(\vep_k -\zeta(T))\;,
\ea
and the elements of the scattering matrix  $P_{ij}$, $i,j=1,\ldots,n$\,:
\ba\label{4} 
P_{ij}(T) &=& \frac{V}{\kBT}  \int\!\di\bk
                              \int\!\di\bk' \, C(\bk,\bk') f^0_k(1-f^0_{k'}) \, u_{ij}(\bk,\bk')\nonumber\\
  u_{ij}(\bk,\bk') &=& [\phi_i(\bk)-\phi_i(\bk')] [\phi_j(\bk) - \phi_j(\bk')],
\ea
where $f^0_k$ is the electronic equilibrium distribution function,
$\bu$ denotes the unit vector in the external field direction,
$\bv$ is the electron velocity,
$\vep_k = (\hbar\bk)^2 / 2m$ -- its energy,
$\zeta(T)$ -- the chemical potential.
$C(\bk,\bk')$ denotes the transition probability per unit time for the free electron scattered from the state $\bk$ to the state $\bk'$.

Considering electron scattering on phonons, nonmagnetic impurities and CF-excitations we will use denotation  $C^{\alpha}(\bk,\bk')$
($\alpha=\mrm{ph,imp,mag}$), and $P_{ij}^{\alpha}(T)$ for the corresponding scattering matrix element (\ref{4}).
In this way, for each $\alpha$ we can specify generalized transport coefficients $L^{\alpha}_{\mu,\nu}$
($\mu,\nu=E,T$) and similarly $\kappa^{\alpha}$, $\sigma^{\alpha}$,
$S^{\alpha}$ -- the $\alpha$ contributions to the thermal conductivity, the electrical conductivity, and the thermoelectric power, respectively.
The expressions for the generalized transport coefficients in the variational approximation of the $n$-th order were derived in Ref.\,[\onlinecite{koh}] with the use of the base functions
\be\label{5} 
\phi_i(\bk) = (\bk\cdot\bu) \left[\vep_{\bk} - \zeta(T)\right]^{i-1}, \qquad i = 1, \ldots, n.
\ee

In the second order, the transport coefficients $L_{\mrm{EE}}^{\alpha}$, $L_{\mrm{TT}}^{\alpha}$, for every $\alpha$ have the form
\ba\label{41} 
L_{\mrm{EE}} &=& \frac{J_1^2}{P_{11}} \left(1+\frac{J_2^2}{J_1^2} \; \frac{P_{11}}{P_{22}} - 2\frac{J_2}{J_1} \; \frac{P_{12}}{P_{22}} \right)\;, \nonumber\\
L_{\mrm{TT}} &=& -\frac1{T} \; \frac{U_2^2}{P_{22}} \left(1+\frac{U_1^2}{U_2^2} \; \frac{P_{22}}{P_{11}} -2\frac{U_1}{U_2}\; \frac{P_{12}}{P_{11}}\right)\;,
\ea
where---for simplicity---we have omitted symbol $\alpha$ in $L^{\alpha}_{\mu,\nu}$
($\mu,\nu=E,T$) and in $P_{ij}^{\alpha}(T)$.

To estimate relations between the components of the sum in (\ref{41}) we use    trial currents relations $J_2/J_1 = \pi^2\,(\kBT)^2/\vepF$, $U_1/U_2=3/(2\vepF)$, following from their form derived with the basis (\ref{5}) $i = 1,2$,  see Eq.\,(9.12.11) in Ref.\,[\onlinecite{zim}]. Taking into account additionally the relations  between the scattering matrix elements, following from their forms (\ref{66}), (\ref{104}), and (\ref{27}) it is easy to verify that, for every $\alpha=\mrm{imp,ph,mag}$, the second and the third components in parentheses in (\ref{41}) are negligible in comparison to the first one.
Hence the following approximations are justified for every $\alpha$
\be\label{44} 
L_{\mrm{EE}} \simeq \frac{J_1^2}{P_{11}}\;, \qquad \ \ L_{\mrm{TT}} \simeq - \frac1{T} \; \frac{U_2^2}{P_{22}}\;,
\ee
where $J_1 = e\,k_F^3 / (3\pi^2\hbar)$,  $U_2 = J_1(\pi\,\kBT)^2 / (3e)$, \cite{zim}, and $k_F$ is the Fermi  radius.

Now we compare the first component of $\kappa$
(\ref{1}), with  the second one $L_{\mrm{TT}}$  (\ref{44})  considering relation $T\,S^2\,L_{\mrm{EE}} / |L_{\mrm{TT}}|$.

For $\alpha=\mrm{mag,imp}$\,, we use the form $S^{\alpha} = -\pi^2\,\kB(\kBT) / (3e\vepF)$, which can be derived, with the scattering matrix elements (\ref{104}) in the  case $\alpha=imp$, and the elements \,(\ref{27}) in the case $\alpha=mag$, in the same way as it has been done in Ref.\,[\onlinecite{asz}] for the conduction electron scattering on the magnetic levels of 4f-electrons in the molecular field.
Substituting in  (\ref{44}) the scattering matrix elements (\ref{104}) and (\ref{27}) we get for the both considered cases
$TL_{\mrm{EE}}\,/ |L_{\mrm{TT}}| = 3e^2 / \pi^2\,\kB^2$ and finally
$TS^2\,L_{\mrm{EE}}\,/ |L_{\mrm{TT}}| = \pi^2(\kBT)^2 / 3(\vepF)^2 \ll 1$.

In  the case of $\alpha=\mrm{ph}$, we use  $(\Sph)^2 < \pi^4\,\kB^2(\kBT)^2/(\vepF)^2$, following from the form $S^{ph}$  in Ref.\,[\onlinecite{zim}],\,(9.12.20).  With the use of the scattering matrix elements (\ref{66}) and (\ref{44})  we get
\be\label{44a} 
\frac{S^2\,T\,L_{\mrm{EE}}}{|L_{\mrm{TT}}|} < \frac{(\kBT)^2}{\vepF^2} \left[ 3\,\pi^2 + \frac{9n_s}{t^2} \right]\equiv R^{\mrm{ph}}\,.
\ee

Because
$R^{\mrm{ph}} \leq (\kBT)^2(3\pi^2 + 9n_s)/\vepF^2$ for $t\geq 1$   and
$R^{\mrm{ph}} \leq (\kB\TD)^2(3\pi^2 + 9n_s)/\vepF^2$ for $t\leq 1$  we obtain
that  $TS^2\,L_{\mrm{EE}}\,/ |L_{TT}| \ll 1$ for every $t$.
 Summarizing,  for every $\alpha$ and all experimentally accessible temperatures, one can neglect the first term in the r.h.s. of (\ref{1}) and approximate
$\kappa^{\alpha}(T) \backsimeq -L_{\mrm{TT}}^{\alpha}$.
Obeying (\ref{44})  we get
\be\label{46} 
\kappa^{\alpha}(T) = \frac{\pi^4\,J_1^2}{9\,e^2} \; \frac{(\kBT)^2}{T\,P_{22}^{\alpha}}.
\ee

Similarly, substituting   $L_{\mrm{EE}}^{\alpha}$ (\ref{44}) into (\ref{1})  we get the $\alpha$ contribution to the electrical resistivity $\rho^{\alpha}=1/\sigma^{\alpha}$
\be\label{47} 
\rho^{\alpha}=\frac{P_{11}^{\alpha}}{J^2_{1}}\,.
\ee

\section{} 
To deal with acoustic phonons we assume Debye model and deformation potential approximation \cite{zim}.
The scattering matrix elements $\Pph_{ij}$ (\ref{4}) derived with the use of the basis (\ref{5}) $i=1,2$  can be found in Refs\,[\onlinecite{koh,koh1,zim}].
In terms of reduced temperature $t=T/\TD$
\begin{equation}
\begin{split} \label{66} 
  \Pph_{11}  =& 4\Pph_0 t^5 \cJ_5(1/t)\,,    \qquad  \qquad
  \Pph_{12}  =  \frac{\veps}2 \Pph_{11}\,,         \\
  \Pph_{22}  =& \frac{\pi^2}3 \,(\kBT)^2\,\Pph_{11}
 \left[\left(1 + \frac3{\pi^2} \frac{n_s}{t^2} \right) - \frac1{2\pi^2} \frac{\cJ_7(1/t)}{\cJ_5(1/t)} \right]\,,
\end{split}
\end{equation}
where
\begin{displaymath}
\int\limits_0^{x} \!\di z \; \frac{z^n}{\sinh^2(z/2)}  \equiv   4 \cJ_n(x)\,,
\end{displaymath}
\\
\noindent are Debye integrals,
$n_s = k_F^2/\qD^2$ is defined by the Fermi $k_F$  and the Debye $\qD$ radii.
$\veps = 2m\,v_s^2$ is the energy of electron of the wave vector $q_s = 2mv_s/\hbar$,
where $v_s$ is the sound velocity averaged over directions in a~crystal.
Value of  $\Pph_0 = 2 C^2 \, V_0 m^2 \qD^5 / (24 \pi^3 \hbar^4 \, M \, v_s)$ is expressed by $C$ -- the energy of electron--phonon interaction,
$V_0$ -- volume of primitive cell,
$m$ -- mass of free electron,
$M$ -- mass of an ion.

Calculating scattering matrix elements $\Pimp_{ij}$ with the  base (\ref{5}) we assumed the standard form of the scattering probability for the conduction electron potential-scattering on the ionized impurities, \cite{ash}
\ba\label{100} 
\Cimp(\bf{k}',\bf{k}) &=& \Cimp_0 \gimp(q) \delta(\vep'-\vep) \;,                            \nonumber\\
\Cimp_0               &=& \frac{32(\pi)^3}V \; \frac{Z^2e^4}{\hbar\,{\epsilon}^2} \; \frac{n^{\mrm{imp}}}{\lambda^4} \;,       \nonumber\\
\gimp(q)              &=& \frac1{(1+(q/\lambda)^2)^2} \,,
\ea
where
$\lambda$ is the screening constant,
$n^{\mrm{imp}}$ -- the density of the impurities,
$Z$ -- the effective valence of the impurity, and
$\epsilon$ -- the dielectric constant.
In the standard approximation\cite{zim},  $\lambda \simeq 8\,k_F / 3$.

Performing the integrals in (\ref{4}) we use identity $f^0(\vepk)(1-f^0(\vep_{\bf{k}})) \equiv \kBT\,\left( -\pt f(\vepk)/\pt \vepk\right)$ and then apply the Sommerfeld expansion (see e.g. Ref.\,[\onlinecite{ash}]) confining to the first non-vanishing term in the approximation of the strong degeneration of the electron gas.

We obtain
\ba\label{104} 
\Pimp_{11} = \Pimp_0 \; \cF(2\,k_F/\lambda)                                   \nonumber\\
\Pimp_{12} = \frac{\pi^2}3 \; (\kBT)^2 \; \frac{\Pimp_{11}}{\vepF}           \nonumber\\
\Pimp_{22} = \frac{\pi^2}3 \; (\kBT)^2 \; \Pimp_{11}
\ea
where
\ba\label{106} 
\cF(p)  &=& \ln(1+p^2) - \frac{p^2}{1+p}\;,                         \nonumber\\
\Pimp_0 &=& \frac4{3\pi} \; \frac{Z^2\,e^4}{\epsilon^2} \; \frac{m^2}{\hbar^5} \; n^{\mrm{imp}}.
\ea

The form (\ref{100}) of scattering probability ensures the existence of the relaxation time solution for the Boltzmann equation.\cite{bra}
That is why, substituting (\ref{104}) in (\ref{46}) and in (\ref{47}) one gets the standard results for the impurity contribution to the thermal conductivity $\kappa^{imp}$, and for the impurity contribution to the electrical resistivity (the residual resistivity $\rhores$) \,\cite{zim}\,\cite{wil}.

\section{} 
The scattering probability for the conduction electron scattered on the CF magnetic excitations in a~cubic crystal has the form\,\cite{hes}
\ba\label{107} 
\Cmag(\bf{k}',\bf{k}) &\!=\!& \Cmag_0 \!\left(\gmag_0 \delta(\vep'\!-\!\vep) +
\!\!\!\sum_{{n=1}\atop{E_m>E_n}}^{2J+1}\! \gmag_{nm}(T) \left[\frac{\delta(\vep'\!-\!\vep\!-\!\Delta_{nm})}{(\exp[z_{nm}]\!-\!1)}
+ \frac{\delta(\vep'\!-\!\vep\!+\!\Delta_{nm})}{(1\!-\!\exp[-z_{nm}])}\right]\right), \nonumber \\
\Cmag_0 &=& \frac{2\pi}{\hbar}\,(\jex(g\!-\!1))^2\,N ,
\ea
where $\jex$ denotes the energy of $s\!-\!f$ exchange interaction, and $g$ is the Lande factor.
The first component in the external brackets refers to the elastic scattering on the zero-frequency CF excitations, the second one to the inelastic scattering on the CF-excitations of the energies
$\pm\Delta_{nm} \equiv \pm(E_m-E_n)\neq 0$   ($z_{nm} = \Delta_{nm}/\kBT$)
and
\ba\label{108} 
&&       g_0  = 3\!\sum_{{n=1}\atop{E_n=E_m}}^{2J+1}\! a_{nm}^{zz} p_n \,,             \nonumber\\
\gmag_{nm}(T) = 3a_{nm}^{zz}(p_{n}&\!-\!&p_{m})\,,                          \qquad \qquad
  a_{nm}^{zz} = |\langle n|J^{z}|m\rangle|^2 ,
\ea
%
where
$J$ denotes the total angular momentum of 4f electrons,
$|n\rangle$, $|m\rangle$ are the 4f eigen-states in the crystalline field and
$E_n$, $E_m$ are the corresponding energies;
$l_n$ is the degeneracy of the level of the energy $E_n$\,, and  $p_n = \exp[-E_n/\kB T]/Z(T)$ is the probability of this level occupation ($Z(T) =\sum_n\,l_n \exp[-E_n/\kB T]$).
$J^{z}$ is the $z$ component of the operator of total angular momentum.
The  matrix elements $a_{nm}^{zz}$ have the 
meaning of the excitations weights.


We will calculate the scattering matrix elements $\Pmag_{ij}$ $i,j=1,2$ (\ref{4}) with the use of the base functions (\ref{5}) in the way described in Ref.\,[\onlinecite{asz}], based on the property of $\Cmag(\bf{k}',\bf{k})$ being even function with $\bf{k}$ and with $\bf{k}'$.
This property, similarly as in the case of electron--impurity scattering ensures the existence of the relaxation time solution of the transport equation\,\cite{bra}, the same for the charge and heat transport.
We write the final form of $\Pmag_{ij}$, $i,j=1,2$ as depending on reduced temperature $t=T/\TD$
\ba\label{27} 
\Pmag_{11}(t)&=&\Pmag_0 \!\sum_{{n=1}\atop{\vep_m>\vep_n}}^{2J+1} \!\!\! 3\,a_{nm}^{zz} p_{n} \frac{1-\exp[-d_{nm}/t]}{\sinh[d_{nm}/t]}
+ \!\!\sum_{{n=1}\atop{\vep_n=\vep_m}}^{2J+1} \!\!\! 3\,a_{nm}^{zz} p_n,                                          \nonumber\\
\Pmag_{12}&=&\frac{\pi^2}{3}(\kBT)^2\,\frac{\Pmag_{11}}{\vepF}                                   \nonumber\\
\Pmag_{22}&=&\frac{\pi^2}{3}(\kBT)^2\,\Pmag_{11}                                                  \nonumber\\
\Pmag_0&=&\frac{2V_{0}j_{ex}^2(g-1)^2}{3\pi^3}\,\frac{m^2\,\kF^4}{\hbar^5},
\ea
where  $\vep_n=E_n/\kB\TD$\,, $\vep_{m}=E_m/\kB\TD$\,, $d_{nm}=\Delta_{nm}/\kB\TD$\,, and
$p_n = \exp[-\vep_n/t]/Z(t)$  with  $Z(t) =\sum_n\,l_n \exp[-\vep_n/t]$.


\begin{flushleft}

\end{flushleft}

\begin{thebibliography}{6}
\bibitem{uw}
{\footnotesize{
It would have followed from the relation  $n_s = (n_a/2)^{2/3}$ (\ref{66}) after substitution $n_a = 3$ (the number of conduction electrons per atom)    corresponding to $RE$In$_3$\,, if there have been one atom per unit cell.
The last condition is necessary for the relation to be fulfilled, within the model we have applied for the electron--phonon scattering\cite{zim}.
Since this condition is not fulfilled in the case of considered here compounds $RE$In$_3$\,,  $n_s$ has not a~strict reference to any measured quantity and we  treat it as a~free parameter.}}
\end{thebibliography}
\end{document}